\documentclass[a4paper, 10pt]{article}

\usepackage[top=1in, bottom=1in, left=1in, right=1in]{geometry}
\usepackage{amsmath}
\usepackage{amssymb}
\usepackage{enumerate}
\usepackage{amsthm}
\usepackage{hyperref}

\begin{document}

\title{\textbf{Comment to a paper of M. Villata on antigravity}}

\author{
        Marcoen J.T.F. Cabbolet\\
        Center for Logic and Philosophy of Science \\
        Vrije Universiteit Brussel\\
        Pleinlaan 2, 1050 Brussels, Belgium\\
        e-mail: Marcoen.Cabbolet@vub.ac.be
        }

\maketitle

\begin{abstract}
In a recent paper of M. Villata, it is claimed that ``antigravity appears as a prediction of general relativity when CPT is applied.'' However, the present paper argues that Villata puts the cart before the horse qua methodology, and that the resulting theory cannot be reconciled with the ontological presuppositions of general relativity. The conclusion is that Villata's suggestion for the physics that might underlie a gravitational repulsion of matter and antimatter is not acceptable in its current state of development.
\end{abstract}

\section{Introduction}
In his recent paper ``CPT symmetry and antimatter gravity in general relativity'', cf. \cite{bib:Vill}, M. Villata extends the paradigm of general relativity (GR) with the assumption of CPT-symmetry. By applying discrete operators for charge, parity and time inversion to the equation of motion in GR, equation (8) in \cite{bib:Vill},
\begin{equation}\label{eq:8}
\frac{\rm{d}^{2}x^{\lambda}}{\rm{d}\tau^{2}}= -\frac{m_{(g)}}{m_{(i)}} \frac{\rm{d}x^{\mu}}{\rm{d}\tau}\Gamma^{\lambda}_{\mu\nu}\frac{\rm{d}x^{\nu}}{\rm{d}\tau}
\tag{8}
\end{equation}
a new equation is constructed, equation (9) in \cite{bib:Vill}:
\begin{equation}\label{eq:9}
\frac{\rm{d}^{2}x^{\lambda}}{\rm{d}\tau^{2}}= -\frac{-m_{(g)}}{m_{(i)}} \frac{\rm{d}x^{\mu}}{\rm{d}\tau}\Gamma^{\lambda}_{\mu\nu}\frac{\rm{d}x^{\nu}}{\rm{d}\tau}
\tag{9}
\end{equation}
This is then interpreted as the equation that governs the motion of antimatter (existing in `our' time-direction) in the gravitational field of ordinary matter; on that basis, Villata claims that ``antigravity appears as a prediction of general relativity when CPT is applied.''\\
 \\
At first glance, this might be a tempting idea to obtain a description of the physics underlying gravitational repulsion. However, to start with, given that quantum physics -- from where the CPT-symmetry is taken -- and relativity theory are two distinct paradigms in physics that are proven to be incompatible, it is epistemologically at least a controversial practise to add a theorem of the one paradigm as an additional assumption to the other. But even if that is ignored, and even if it is assumed that the derivation of (9) from (8) is mathematically correct, the next section argues that this method of theory construction is \emph{in itself} inadmissible, and that the theory that results from adding eq. (9) to GR cannot be reconciled with the ontological presuppositions of GR. The final section discusses the implications thereof.

\section{Shortcomings of Villata's method and result}
CPT-symmetry is a law at the metalevel that follows from the actual laws of physics at object level. In other words, from the theory of what gravitation actually is it should be clear at object level what the process of gravitational interaction for matter is and what the process of gravitational interaction for antimatter is, and from there it should follow at the metalevel that CPT-symmetry holds (or doesn't hold) between these processes. In theory development, it is one thing to assume a symmetry as a condition that has to be satisfied by a yet to be developed theory, but Villata puts the cart before the horse: he assumes CPT-symmetry and uses the operators C, T, and P \emph{as if} these are applicable to derive the theory of what the process of gravitational interaction of antimatter is at object level from the theory of the corresponding process of matter. But these operators cannot be applied that way: this method of theory development is inadmissible. \\
 \\
This inadequateness of the method comes to expression in the fact that the theory that results from adding equation (9) to GR, cannot be reconciled with the ontological presuppositions of GR. In GR, namely, a classical ontology of particles and fields is presupposed: a particle is an object of negligible dimensions that at every point in time has a definite position, a definite spatial momentum, and possibly a nonzero rest mass; the gravitational field is nothing but the metric of spacetime, where spacetime is the set of \emph{all} events. That Villata's theory cannot be reconciled with this ontology is then best demonstrated by a Gedanken-experiment.

Consider the gravitational field of the earth, and consider the event that a neutron with initial spatial velocity $\underline{v}_1$ is created at spatiotemporal position $\langle t_1, \underline{x}_1 \rangle$ by the production of a particle/antiparticle pair. The neutron ``sees" the gravitational field of the earth: the initial position and velocity then completely determine its further trajectory through the gravitational field of the earth. Now consider the gravitational field of the earth to be constant, and consider that in a second event by the production of another particle-antiparticle pair an antineutron has been created at spatiotemporal position $\langle t_2, \underline{x}_2 \rangle$  with spatial velocity $\underline{v}_2$ and with $\underline{x}_2 = \underline{x}_1$ and  $\underline{v}_2 = \underline{v}_1$. Given that spacetime in GR is the set of \emph{all} events, this second event thus takes place \emph{in that same spacetime}. In the physical picture of GR, this antineutron is then nothing but another particle that exists in the one and only spacetime, and that ``sees" the same gravitational field as the neutron earlier at $t = t_1$: there is nothing more to it -- on the basis of the presupposed ontology of GR one would, thus, expect that the antineutron follows the same spatial trajectory as the neutron. In [1], however, it is claimed that eq. (9) \emph{nevertheless} dictates that antineutrons and neutrons behave differently in the gravitational field of the earth. The crux is then that the difference between the neutron and the antineutron is thereby nothing but the assumed C-inversion\footnote{The other two operators, time inversion T and parity inversion P, apply to an ontology of \emph{processes}, not \emph{particles}!}: the neutron and the antineutron differ, thus, only with respect to some quantum numbers that have no relevance for gravitation whatsoever. So, given that the antineutron necessarily exists in the same spacetime as the neutron and ``sees'' the same gravitational field, this C-inversion can thus not possibly underlie a difference in behavior under the influence of gravitation.

This demonstrates that the theory, obtained by adding the derived eq. (9) to GR, cannot be reconciled with the ontological presuppositions and the physical picture of GR.

\section{Implications}
The main implication is that Villata's description of a gravitational repulsion of matter and antimatter is not acceptable in its current state: the main point is that in the framework of GR, spacetime is the set of \emph{all} events -- the creation of an antiparticle (an event) happens thus in that one spacetime. That is to say: the existence of an inverted spacetime has first to be assumed before it can be said that antiparticles exist (or ``live'') in an inverted spacetime. Thus, if gravitational repulsion of matter and antimatter were a fact of nature, then there are only two possibilities:
\begin{enumerate}
\item	GR is correct for ordinary particles, but the ontology has to be extended so that antiparticles ``see'' a different metric than ordinary particles;
\item	the entire ontology of GR is at fault, and gravitation in reality is something completely different.
\end{enumerate}
The first possibility has already been investigated by Santilli; the result is published in \cite{bib:Sant}. If Villata were to adjust his theory ontologically such that antimatter exists in an inverted spacetime, then this requires an additional discussion on how this differs from Santilli's theory, as well as a discussion about photons -- a photon is identical to its antiparticle, yet it sees the same metric as ordinary matter. It should be noted, however, that this classical approach to a description of gravitational repulsion will not yield new physics that provide a new point of view to the unification of GR and quantum mechanics (QM): it will thus not bring a solution to this main problem of contemporary physics any closer.

The second possibility has also been investigated already. In the study, published in \cite{bib:Cabb}, a completely new ontology has been identified: a formal axiomatic system with a physical interpretation has been defined, so that indivisible components of the universe are referred to by individual constants of the formal language (formalism), and so that fundamental principles are rigorously formalized as relations between components. These fundamental principles (that are nonlogical axioms of the axiomatic system) are thus built from very different elements than the axioms of GR or QM, but together these principles from a physically complete set, called the Elementary Process Theory (EPT). This is not an interaction theory, but a theory that describes the dynamics of the individual processes: these are essentially all the same in the universe of the EPT. It has been demonstrated how gravitational repulsion functions in this universe: contrary to the aforementioned classical apporoach to a description of gravitational repulsion, this new approach does provide a new point of view to the description of gravitation and electromagnetism as two aspects of a single long-distance interaction.\\
 \\
Villata's suggestion in \cite{bib:Vill} that a repulsive interaction of matter and antimatter might underly the observed expansion of the universe remains worthy of further investigation. To this can be added that the new ontology presented in \cite{bib:Cabb} forms the basis for a different approach to this problem; another question is then whether cosmological dark matter can be described with this ontology.


\begin{thebibliography}{10}
\bibitem{bib:Vill}
{\sc Villata},~M., CPT symmetry and antimatter gravity in general relativity, \emph{EPL} \textbf{94}(2), 20001 (2011);
\href{http://arxiv.org/abs/1103.4937}{arXiv:1103.4937 [gr-qc]}
\bibitem{bib:Sant}
{\sc Santilli},~R.M., A classical isodual theory of antimatter and its prediction of antigravity, \emph{Int. J. Mod. Ph. A} \textbf{14}(14), 2205-2238 (1999);
\href{http://www.worldscinet.com/ijmpa/14/1414/S0217751X99001111.html}{DOI: 10.1142/S0217751X99001111}
\bibitem{bib:Cabb}
{\sc Cabbolet},~M.J.T.F., Elementary Process Theory: a formal axiomatic system with a potential application as a foundational framework for physics supporting gravitational repulsion of matter and antimatter, \emph{Ann. Phys. (Berlin)} \textbf{522}(10), 699-738 (2010);
\href{http://onlinelibrary.wiley.com/doi/10.1002/andp.201000063/abstract}{DOI: 10.1002/andp.201000063}
\end{thebibliography}
\end{document}